\documentclass[prd,preprint,tightenlines,floatfix,showpacs,preprintnumbers,nofootinbib,eqsecnum,superscriptaddress]{revtex4}

 \usepackage[dvips,final]{graphicx}
  \usepackage{amssymb}
   \usepackage{amsmath}
    \usepackage{amsfonts}
     \usepackage{epsfig}
      \usepackage{bm}

\usepackage[section]{placeins}

\usepackage{multirow}
\usepackage{ctable}
\usepackage{booktabs}
\usepackage{array}
\usepackage{tabularx}
\usepackage{xcolor}
\usepackage{pstricks}

\def\lsim{\mathrel{\rlap{\lower4pt\hbox{\hskip1pt$\sim$}}
    \raise1pt\hbox{$<$}}}         
\def\gsim{\mathrel{\rlap{\lower4pt\hbox{\hskip1pt$\sim$}}
    \raise1pt\hbox{$>$}}}         

\begin{document}
\title{Double scattering production of two positron-electron pairs\\
in ultraperipheral heavy-ion collisions}

\author{Mariola K{\l}usek-Gawenda}
\email{mariola.klusek@ifj.edu.pl} \affiliation{Institute of Nuclear
Physics PAN, PL-31-342 Cracow,
Poland}

\author{Antoni Szczurek}
\email{antoni.szczurek@ifj.edu.pl} 
\affiliation{Institute of Nuclear
Physics PAN, PL-31-342 Cracow,
Poland}

\affiliation{
University of Rzesz\'ow, PL-35-959 Rzesz\'ow, Poland}

\date{\today}

\begin{abstract}
We present first measurable predictions for electromagnetic (two-photon) 
double scattering production of two positron-electron pairs 
in ultraperipheral heavy-ion collisions at LHC.
Measureable cross sections are obtained with realistic cuts
on electron/positron (pseudo)rapidities and transverse momenta
for the ALICE and ATLAS or CMS experiments.
The predictions for total and differential cross sections
are presented. We show also two-dimensional distributions
in rapidities of the opposite-sign (from the same or different subcollisions)
and of the same-sign ($e^+ e^+$ or $e^- e^-$) electrons
and in rapidity distance between them.
Expected number of events are presented and discussed.
Our calculations strongly suggest that relevant measurements
with the help of ATLAS, CMS and ALICE detectors
are possible in a near future.

\end{abstract}


\maketitle

\section{Introduction}

Multiple scattering effects are present in many reactions
at high energies such as proton-nucleus (multiple nucleon-nucleon
scatterings), proton-proton (double parton scatterings) and 
in ultraperipheral collisions (UPC) of heavy-ions.
The double parton scattering effects in proton-proton collisions 
become increasingly important
with steadily increasing energy in proton-proton collisions
\cite{MPI_proceedings}.
The cross section for double (multiple) scattering can be large 
provided the cross section for single parton scattering is large.
The best example is double charm production 
in proton-proton collisions \cite{ccbar_ifj}.

Not much attention was devoted to multiple scattering in UPC of heavy ions.
In UPC of heavy ions, where in real experiments the integrated
luminosity is rather small, in our opinion, 
only cross section for $A A \to A A \rho^0$ 
and $A A \to A A e^+ e^-$ reactions are large enough \cite{UPC_reviews}
to potentially observe double scattering effects.
The double-scattering (DS) mechanism for $\rho^0 \rho^0$ production
was studied e.g. in \cite{Klein-Nystrand,KS2014}.
So far our prediction for four charged pion production was confronted
only with the STAR data \cite{STAR_4pi}. There the DS mechanism
was not sufficient \cite{KS2014} to explain the existing STAR data 
\cite{STAR_4pi}.
Production of other meson combinations was discussed very recently
in \cite{GMN2016}.
 
\begin{figure}[!h]
\includegraphics[scale=0.3]{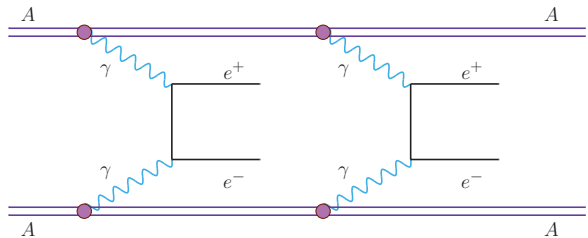}
\caption{Double-scattering mechanism for $e^+e^-e^+e^-$ production
in ultrarelativistic UPC of heavy ions.}
\label{fig:AA_AA4e}
\end{figure} 
 
The double production of two dielectron pairs (see Fig.~\ref{fig:AA_AA4e}) was discussed
e.g. in the context of bound-free production \cite{Serbo}.
There rather total cross section is discussed.
The total cross section is dominated by the very low transverse momenta of electrons. 
The low-transverse momentum electrons cannot be, however,
measured neither at the LHC. Here we wish to 
make first predictions that have a chance to 
be verified experimentally at the LHC.
Such a measurement would allow to verify (for the first time) our 
understanding of the underlying double scattering reaction mechanism
in ultraperipheral heavy-ion collisions. We wish to emphasize
that so far no double scattering mechanism in UPC was confirmed or
unambigously verified by experimental results on UPC of heavy ions. 
As we will show in the following the 
$Pb Pb \to Pb Pb e^+ e^- e^+ e^-$ (see Fig.~\ref{fig:AA_AA4e}) is a good candidate 
which has a good chance to be the first case in this context.

\section{Sketch of the formalism}

The cross section for single $e^+ e^-$ production is calculated as
described in Ref.\cite{KS_mumu}. The total cross section
can be written as: 
\begin{eqnarray}
\sigma_{A_1 A_2 \to A_1 A_2 e^+e^-}\left(\sqrt{s_{A_1A_2}} \right) &=& \int \sigma_{\gamma \gamma \to e^+e^-} 
\left(W_{\gamma\gamma} \right)
N\left(\omega_1, {\bf b_1} \right)
N\left(\omega_2, {\bf b_2} \right) 
S_{abs}^2\left({\bf b}\right) \nonumber \\
& \times &
2\pi b \mathrm{d} b \, \mathrm{d}\overline{b}_x \, \mathrm{d}\overline{b}_y \, 
\frac{W_{\gamma\gamma}}{2}
\mathrm{d} W_{\gamma\gamma} \, \mathrm{d} Y_{e^+e^-}  \;,
\end{eqnarray}
where $N(\omega_i,{\bf b_i})$ are photon fluxes,
$W_{\gamma\gamma}=M_{e^+e^-}$
and $Y_{e^+e^-}=\left( y_{e^+} + y_{e^-} \right)/2$ 
is a invariant mass and a rapidity of the outgoing $e^+e^-$ system,
respectively. 
Energy of photons is expressed through $\omega_{1/2} = W_{\gamma\gamma}/2 \exp(\pm Y_{e^+e^-})$.
$\bf b_1$ and $\bf b_2$ are impact parameters 
of the photon-photon collision point with respect to parent
nuclei 1 and 2, respectively, 
and ${\bf b} = {\bf b_1} - {\bf b_2}$ is the standard impact parameter 
for the $A_1 A_2$ collision. The quantities 
$\overline{b}_x$ and $\overline{b}_y$  are the components of the $(\bf b_1 + \bf b_2)/2$:
$\overline{b}_x = (b_{1x}+b_{2x})/2$ and $\overline{b}_y = (b_{1y}+b_{2y})/2$.
The five-fold integration is 
performed numerically.
For more details see \cite{KS_mumu}.
Only in approximate case of simplified
charge form factor the integration can be done analytically \cite{UPC_reviews}.
In both cases the integrated cross section can be then written formally as 
\begin{equation}
\sigma_{A_1 A_2 \to A_1 A_2 e^+e^-}= \int P_{\gamma \gamma \to e^+e^-}(b)\, \mathrm{d}^2 b \; .
\label{b_integration}
\end{equation}
Here $P_{\gamma \gamma \to e^+e^-}(b)$ 
has an interpretation of a probability to produce
a single $e^+ e^-$ pair in the collision at the impact parameter $b$.
This general formula is not very useful for practical calculation
of double scattering.
If the calculation is done naively $P_{\gamma \gamma \to e^+e^-}(b)$ can be larger than 1
in the region of low impact parameter.
Then a unitarization procedure is needed \cite{Serbo_unitarity}.

If one wishes to impose some cuts on produced particles (electron, positron) which come from
experimental requirements or to have distribution in some helpful and interesting kinematical
variables of individual particles (here $e^+$ or $e^-$), 
more complicated calculations are required \cite{KGLSz}. 
To have detailed information about rapidities of individual electrons 
an extra integration over a kinematical variable describing 
angular distribution for the $\gamma\gamma \to e^+e^-$ subprocess is required
and the total $\sigma_{\gamma\gamma \to e^+e^-}$ cross section
has to be replaced by relevant differential cross section.
Then formula (\ref{b_integration}) can be written more differentially 
in kinematical variables of the produced leptons (rapidities and
transverse momenta) as:
\begin{equation}
\frac{d \sigma_{A_1 A_2 \to A_1 A_2 e^+e^-}}{d y_{+} d y_{-} d p_t} =
\int \frac{\mathrm{d} P_{\gamma \gamma \to e^+e^-}(b;y_{+},y_{-},p_t)}{d y_{+} d y_{-} d p_t} \, \mathrm{d}^2 b \; .
\label{b_integration_differentially}
\end{equation}
Other choices of kinematical variables are possible as well.
If one imposes cuts on transverse momenta of leptons the probabilities
becoming small and no unitarization is needed.

The cross section for double scattering can be then 
written as:
\begin{eqnarray}
\frac{\mathrm{d} \sigma_{AA \to AA e^+e^-e^+e^-}}{\mathrm{d}y_1 \mathrm{d}y_2 \mathrm{d}y_3 \mathrm{d}y_4} &=& \frac{1}{2} 
\int  	\left( 	\frac{\mathrm{d} P_{\gamma \gamma \to e^+e^-} \left( b,y_1,y_2;p_t > p_{t,cut} \right)}{\mathrm{d}y_1 \mathrm{d}y_2}
\times               \frac{\mathrm{d} P_{\gamma \gamma \to e^+e^-} \left( b,y_3,y_4;p_t > p_{t,cut} \right)}{\mathrm{d}y_3 \mathrm{d}y_4} \right) \nonumber  \\ 
& \times & 2 \pi b \, \mathrm{d} b  \;.
\end{eqnarray}
The combinatorial factor $1/2$ takes into account identity of the two pairs.
We shall use the formula above to estimate the double
scattering cross sections.

In our calculations here we use both realistic fluxes of photons calculated
with charge form factors of a nucleus, being Fourier transform of realistic
charge distributions or a more simplified formula
from \cite{KS_mumu} is used.

From the technical point of view, first
$\frac{d P_{\gamma \gamma \to e^+ e^-}(b,y_1,y_2;p_t 
> p_{t,cut})}{d y_1 d y_2}$ are calculated
on the three-dimensional grid in $b$, $y_1$ and $y_2$.
Then in the next step those grids are used to calculate
the cross sections corresponding to double scattering.
We use the MC-based numerical integration program VEGAS \cite{MC_Lepage}.
For test we use also a grid-type integration.

\section{First results}

\begin{figure}[h!]
\includegraphics[scale=0.35]{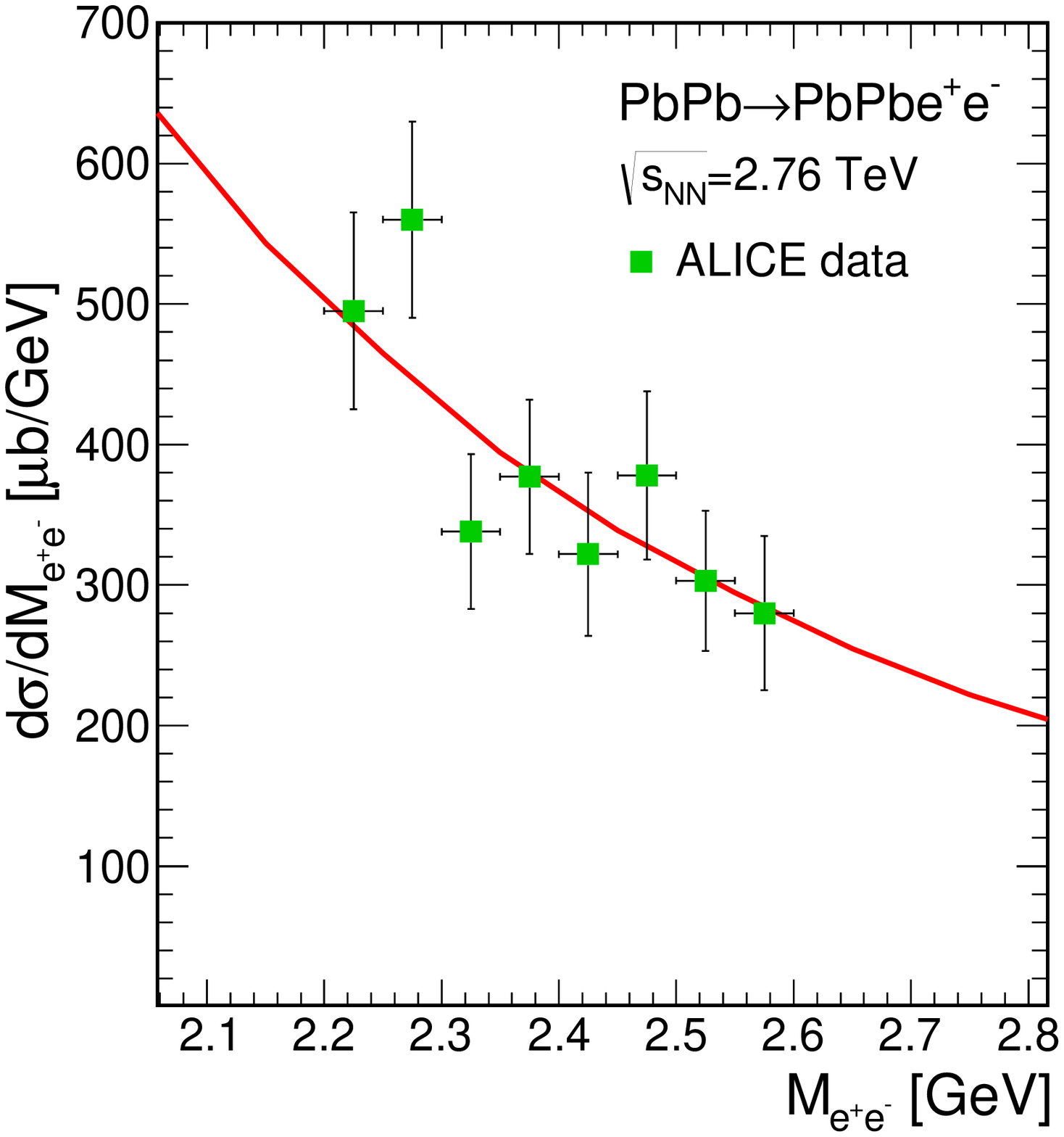}
\includegraphics[scale=0.35]{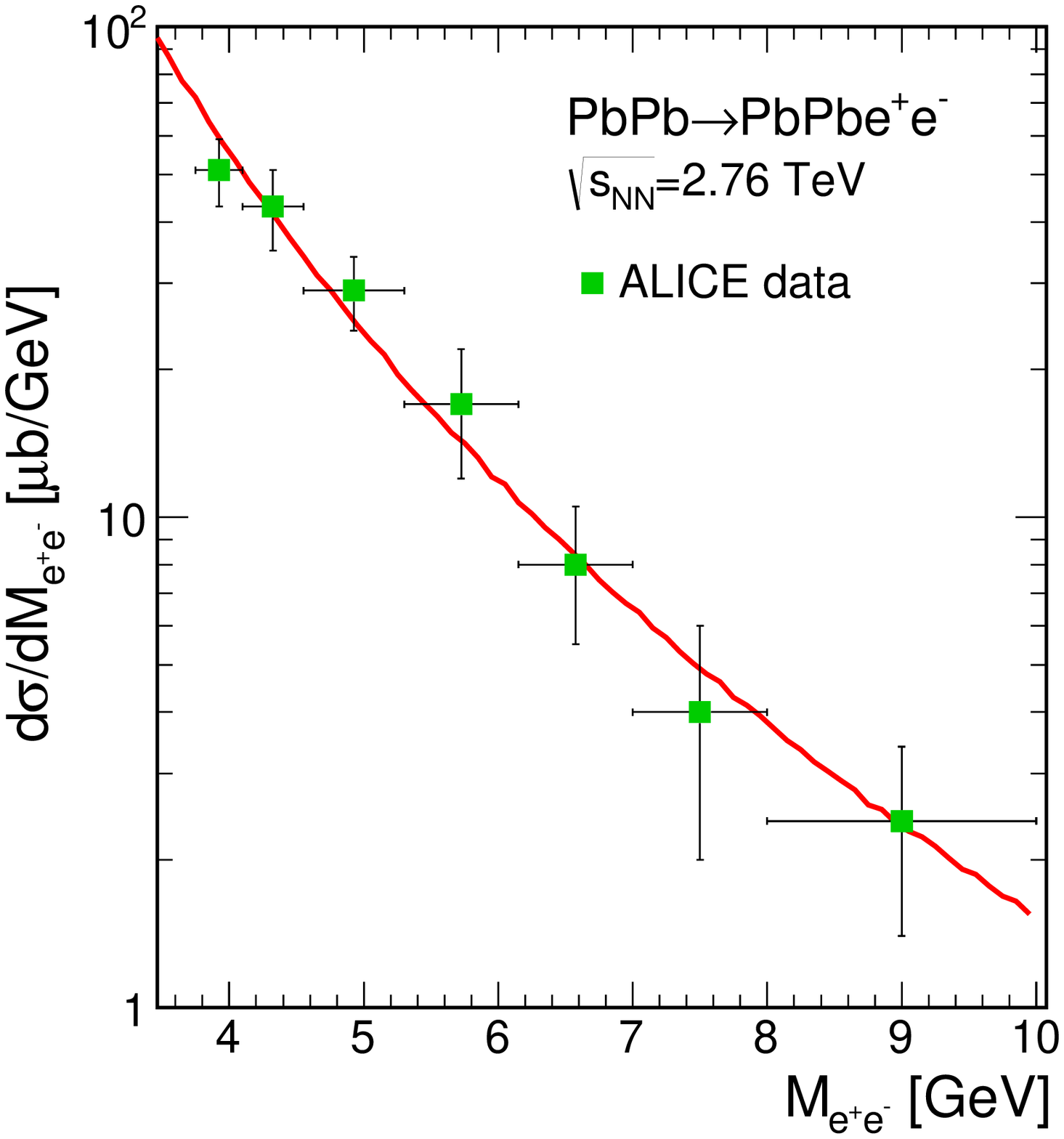}
\caption{
Invariant mass distributions of dielectrons in UPC of heavy ions calculated
within our approach \cite{KS_mumu} together with the recent ALICE data 
\cite{ALICE_epem}.}
\label{fig:dsig_dMee_ALICE}
\end{figure}

Before we present our results for $e^+ e^- e^+ e^-$ production 
let us compare our results with existing experimental data 
for single $e^+ e^-$ pair production.
In Fig.~\ref{fig:dsig_dMee_ALICE} our results are compared 
with recent ALICE data \cite{ALICE_epem}.
Here we consider lead-lead UPC at $\sqrt{s_{NN}}=2.76$ TeV with $|y_e|<0.9$.
The left panel shows the ALICE data \cite{ALICE_epem} for 2.2 GeV $<M_{ee}<$ 2.6 GeV and
the right panel shows their results for 3.7 GeV $<M_{ee}<$ 10 GeV.
Our results for single-scattering mechanism 
almost coincide with the experimental data.

\begin{figure}
\includegraphics[scale=0.35]{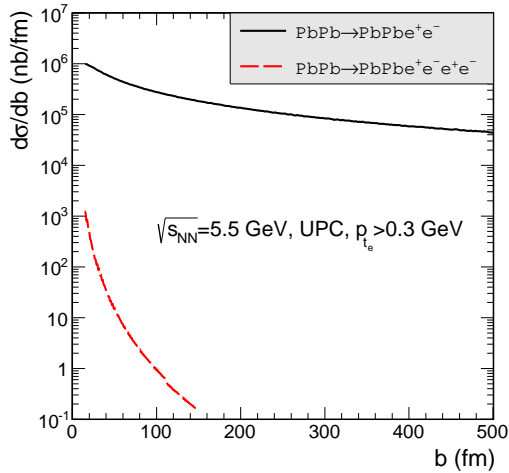}
\caption{Differential cross section as a function of
impact parameter (distance between two colliding nuclei).
The upper line denotes result for the PbPb$\to$PbPb$e^+e^-$ reaction and 
the lower line shows result for the PbPb$\to$PbPb$e^+e^-e^+e^-$ reaction.}
\label{fig:dsig_db}
\end{figure}

Having shown that our approach allows to describe single pair production
we can go to our predictions for two $e^+ e^-$ pair production.
Now we are going to discuss briefly a purely theoretical distribution.
Fig.~\ref{fig:dsig_db} shows differential cross section as a function
of impact parameter (distance between two nuclei) 
for lead-lead UPC at $\sqrt{s_{NN}}=5.5$ TeV 
and $p_{t,e}>0.3$ GeV. 
One can see that the cross section for $e^+e^-e^+e^-$ production
drops off much faster than in the case of single $e^+e^-$ production.
The probability for the production of four particles is of course
much lower than the probability for production of one electron-positron pair.

In Table \ref{tab:list} we have collected integrated cross sections for different
experimental cuts corresponding to ALICE and ATLAS or CMS experiments.
In the ATLAS case we show result for the tracking detectors ($|\eta|<2.5$)
as well as including forward calorimeters ($|\eta|<4.9$).
The rapidity coverage of the CMS calorimeters is very similar.
In the later case
particle identification (PID) is much worse than for the tracker.
However, the cross sections are then much larger than when using
the tracker only.
The main ALICE detector allows for the particle identification
practically down to transverse momenta of 
0.2 GeV, which makes it rather special.
The number for full rapidity coverage and $p_t>$ 0.2 GeV given in the table 
is much (three orders of magnitude)
smaller than the total cross section for two pair production \cite{Serbo},
where it was estimated to be about 10 mb.

\begin{table}[h!]
\caption{Nuclear cross section for $PbPb \to PbPb e^+e^-e^+e^-$ at $\sqrt{s_{NN}}=5.5$ TeV
for different cuts specified in the table.}
\begin{tabular}{|l|r|r|}
\hline
cut set					& $\sigma_{UPC}$		& N$_{\mbox{events}}$ for L=1 nb$^{-1}$ 		\\ \hline \hline
$p_{t_{e}}>0.2$ GeV		& 52.525 $\mu$b 		& 52 525		\\ 
$p_{t_{e}}>0.2$ GeV, 
$|y_e|<2.5$				& 10.636 $\mu$b 		& 10 636 	\\
$p_{t_{e}}>0.2$ GeV, 
$|y_e|<1$				& 0.649 $\mu$b 		& 649		\\ \hline
$p_{t_{e}}>0.3$ GeV,
$|y_e|<4.9$				& 7.447 $\mu$b 		& 7 447		\\ 
$p_{t_{e}}>0.3$ GeV,
$|y_e|<2.5$				& 2.052 $\mu$b		& 2 052		\\	\hline
$p_{t_{e}}>0.5$ GeV,
$|y_e|<4.9$				& 0.704 $\mu$b		& 704		\\ 
$p_{t_{e}}>0.5$ GeV,
$|y_e|<2.5$				& 0.235 $\mu$b		& 235		\\ \hline
$p_{t_{e}}>1$ GeV		& 25.2 nb			& 25			\\ 
$p_{t_{e}}>1$ GeV,
$|y_e|<4.9$				& 22.6 nb  			& 23			\\ 
$p_{t_{e}}>1$ GeV,
$|y_e|<2.5$				& 9.8 nb				& 10 		\\
$p_{t_{e}}>1$ GeV, 
$|y_e|<1$				& 0.6 nb				& 1			\\	
\hline
\end{tabular}
\label{tab:list}
\end{table}

\begin{figure}
\includegraphics[scale=0.3]{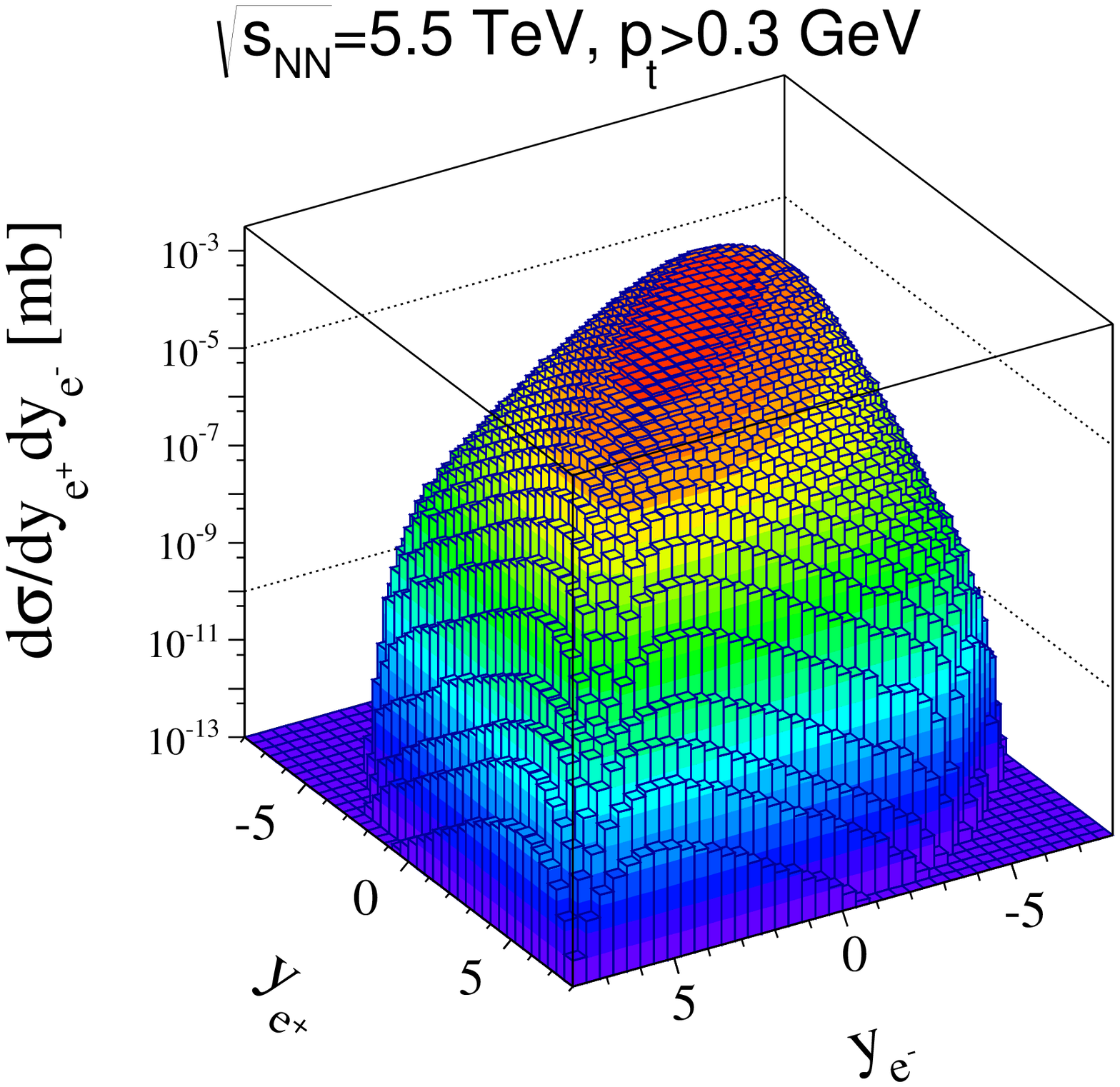}
\includegraphics[scale=0.3]{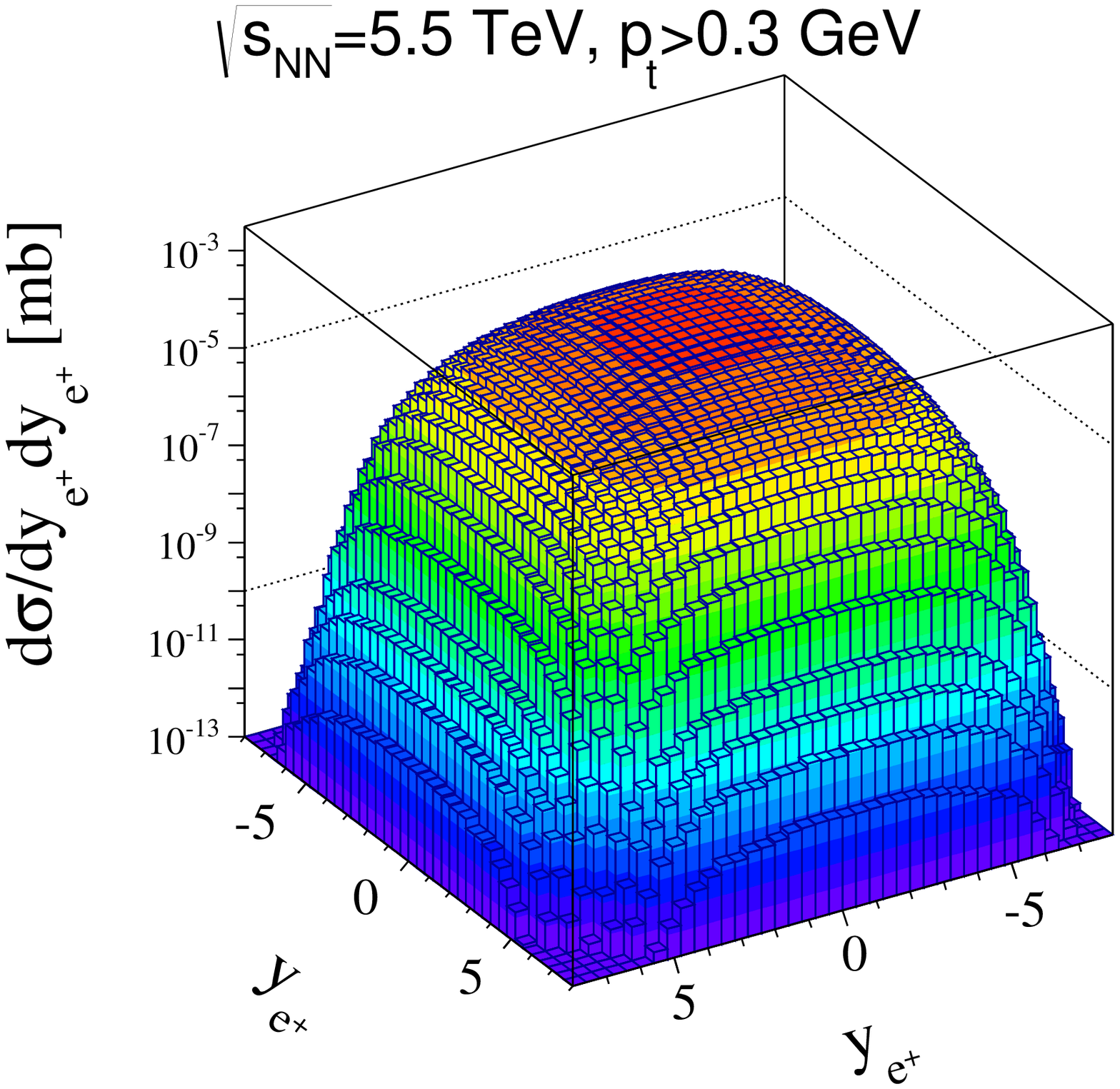}
\caption{Two-dimensional distribution in rapidities of the opposite-sign
leptons from the same collision (left panel) and for the same-sign leptons (right
panel). The cross section for the $e^+e^-e^+e^-$ production
is calculated for lead-lead UPC at $\sqrt{s_{NN}}=5.5$ TeV
and $p_t>0.3$ GeV.
}
\label{fig:dsig_dy1dy2}
\end{figure}

In Fig.\ref{fig:dsig_dy1dy2} we show our predictions 
for the opossite-sign $d\sigma/dy_1dy_2$ (left panel) 
and the same-sign $d\sigma/dy_1dy_3$ (right panel) electrons.
We omit here trivial (experimental) factor 2 
(two possibilities: two-scatterings for opposite sign and 
two signs of electrons for the same sign case).
While the $e^+$ and $e^-$ are correlated by the matrix element
for the $\gamma \gamma \to e^+ e^-$ subprocess the 
$e^+ e^+$ (or $e^- e^-$) are 
not correlated. As a consequence the two-dimensional distributions 
in rapidities are broader for the case of the same-sign electrons.

\begin{figure}
\includegraphics[scale=0.3]{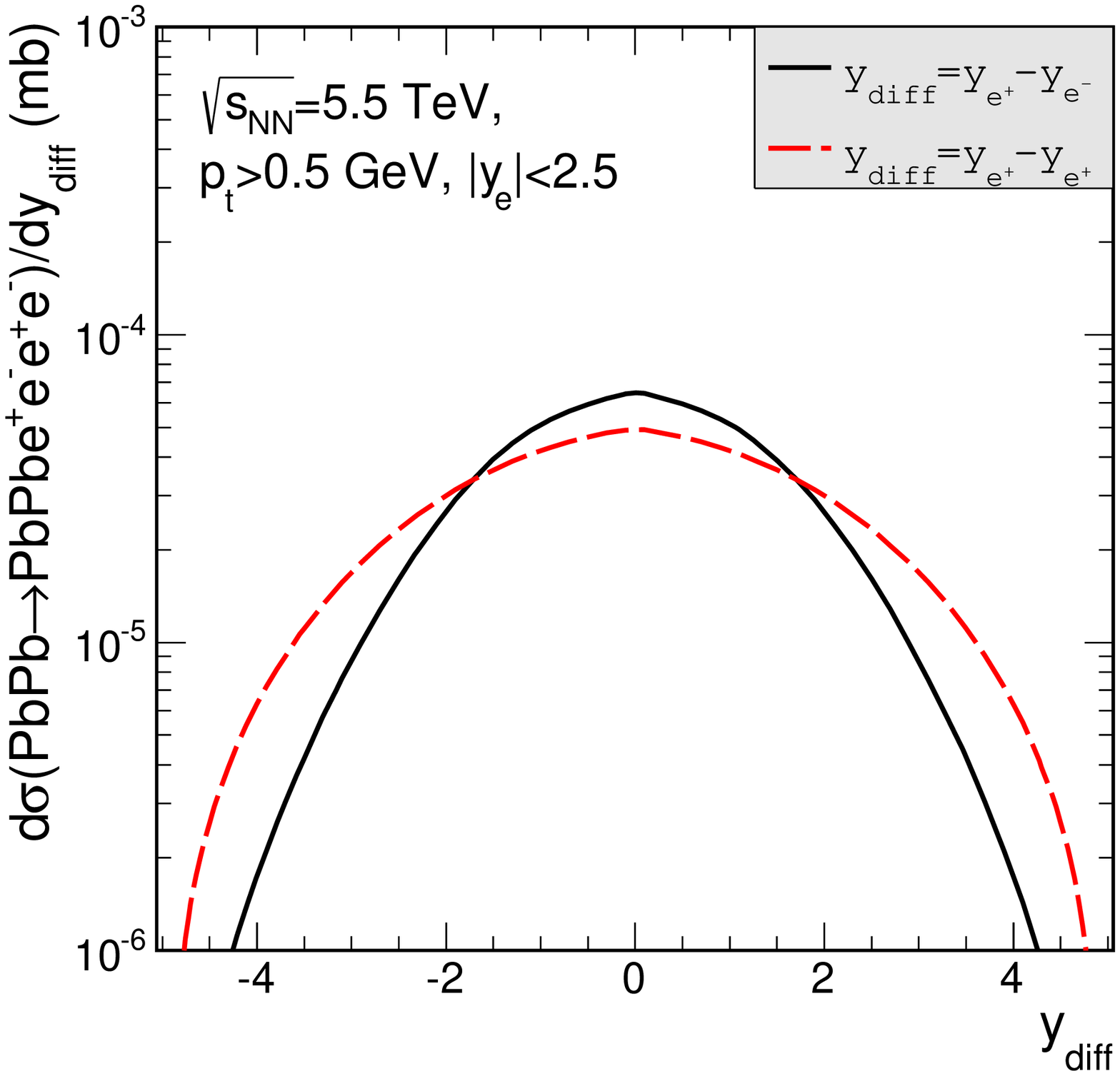}
\includegraphics[scale=0.3]{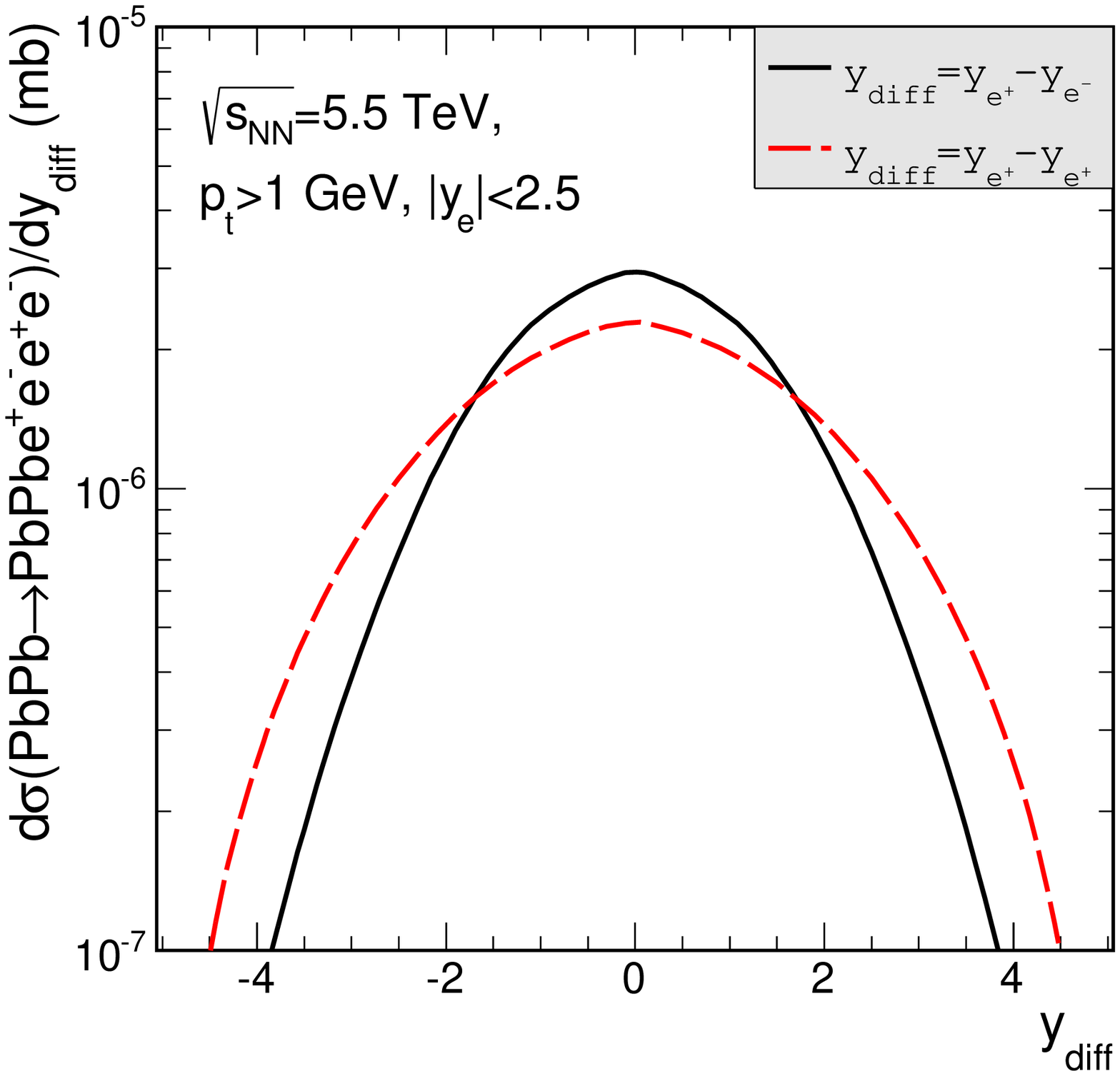}
\caption{Distributions in rapidity difference between the opposite-sign
electrons (solid line) and between the same-sign electrons (or positrons) 
from the same subcollision (dashed line)
for two different lower cuts on lepton transverse momenta: 0.5 GeV (left
panel) and 1.0 GeV (right panel). This calculation is done assuming
that electrons/positrons are measured by the ATLAS main tracker.
}
\label{fig:dsig_dydiff}
\end{figure}

In Fig.\ref{fig:dsig_dydiff} we compare results for
$d \sigma/d y_{diff}$ as a function of rapidity difference between
the same-sign (solid line) and, from the same subcollision opposite-sign 
(dashed line) electrons
assuming each of the electrons/positrons to be
within the ATLAS main detector (-2.5 $<\eta_+, \eta_{-} <$ 2.5)
for transverse momenta $p_t >$ 0.5 GeV (left panel) and for 
$p_t >$ 1 GeV (right panel). 
Such distributions can, in our opinion, be measured at the LHC
and could allow for a first verification of the double scattering
mechanism in UPC of heavy ions. We wish to remind here that such a verification was
not possible for the double scattering production of two $\rho^0$ mesons
\cite{KS2014} where other, at the moment not well understood, mechanisms 
probably play the dominant role \cite{KS2014}.

\begin{figure}
\includegraphics[scale=0.3]{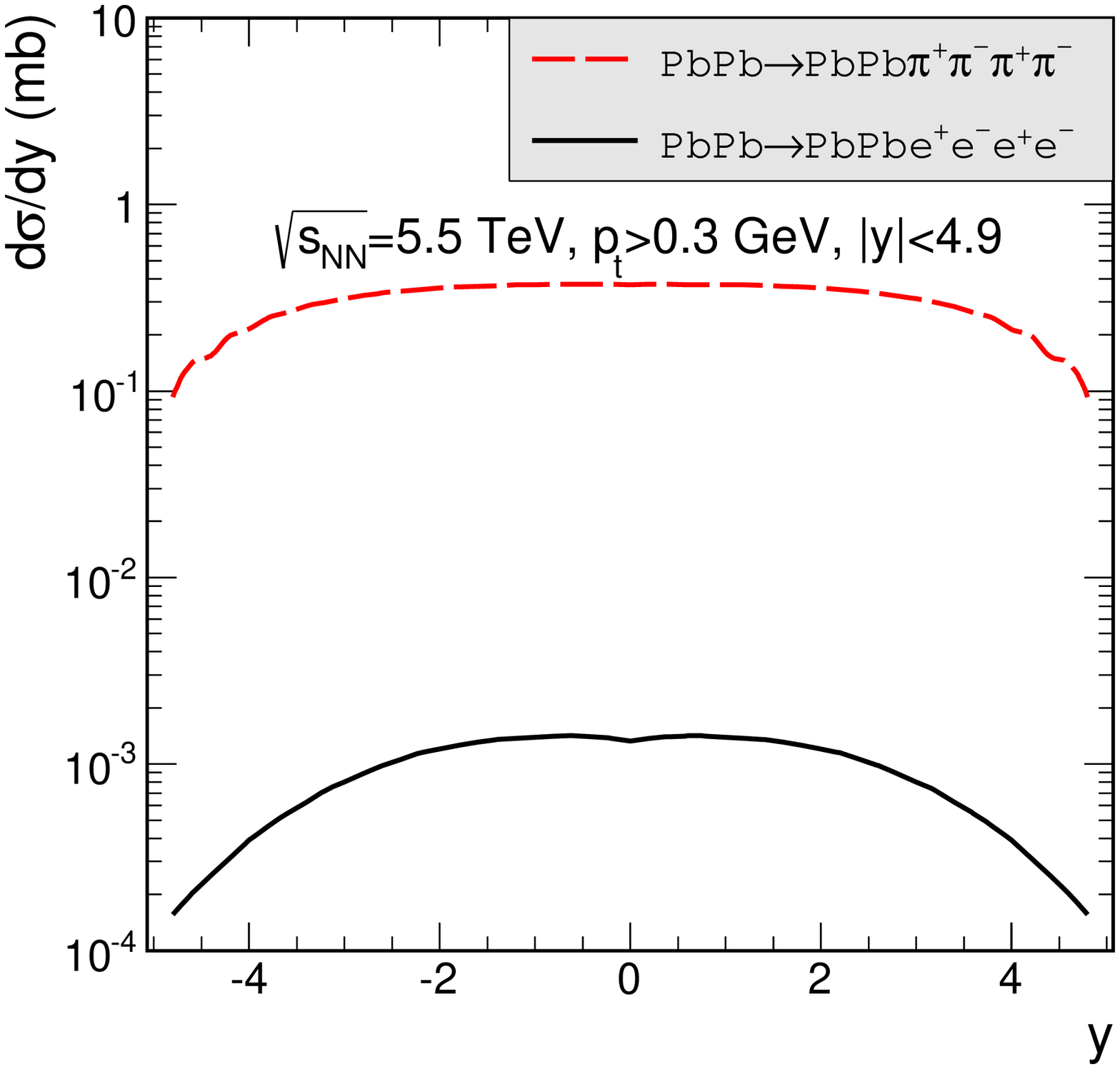}
\includegraphics[scale=0.3]{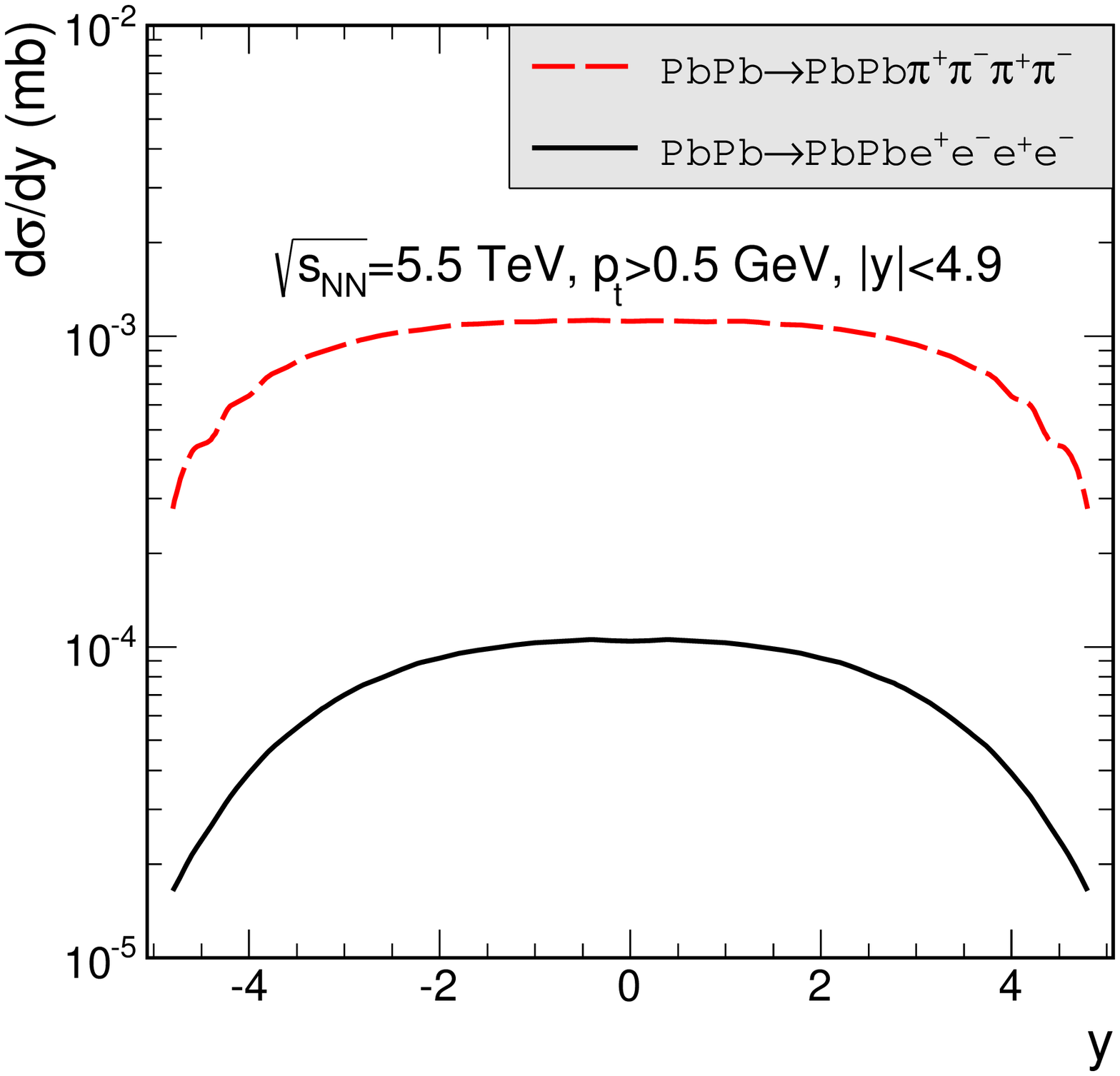}
\caption{Rapidity distribution of electron/positron (solid line)
and charged pion (dashed line) for lead-lead collisions at the LHC ($\sqrt{s_{NN}}=5.5$ TeV)
together with the limitation on rapidity and transverse momenta of each single outgoing particle.
The left panel shows results with limitation on $p_t>$ 0.3 GeV and the right panel
corresponds to $p_{t}>$ 0.5 GeV.}
\label{fig:dsig_dy_e_pi}
\end{figure}

Finally we wish to discuss briefly potential background(s).
The PbPb$\to$PbPb$\pi^+\pi^-\pi^+\pi^-$ reaction discussed in Ref. \cite{KS2014}
is a possibility.
Here we include only double scattering production of two $\rho^0$ mesons
which decay into four pions. In Fig.~\ref{fig:dsig_dy_e_pi} we show
a comparison of the cross sections for the $e^+e^-e^+e^-$ and $\pi^+\pi^-\pi^+\pi^-$
final states for two different lower cuts on transverse momenta.
The cross section for four pions is much bigger than the cross section
for four electrons.
The situation improves when increasing the lower cut.

\begin{table}[h!]
\caption{Nuclear cross section for the $PbPb \to PbPb \pi^+\pi^-\pi^+\pi^-$ and 
$PbPb \to PbPb e^+e^-e^+e^-$ reactions at $\sqrt{s_{NN}}=5.5$ TeV
with $|y|<$ 4.9 and for different cuts on transverse momenta of pions or electrons.}
\begin{tabular}{|l|r|r|}
\hline
Reaction									& $p_{t,min}	$ = 0.3 GeV 	& $p_{t,min}	$ = 0.5 GeV \\ \hline \hline	
$PbPb \to PbPb \pi^+\pi^-\pi^+\pi^-$ 	& 2.954 mb 				& 8.862 $\mu$b			\\	
$PbPb \to PbPb e^+e^-e^+e^-$ 			& 7.447 $\mu$b			&  0.704 $\mu$b			\\ \hline
\end{tabular}
\label{tab:4pi_4e}
\end{table}

In Table \ref{tab:4pi_4e} we show the cross section for the signal 
($e^+e^-e^+e^-$) and the reducible background ($\pi^+\pi^-\pi^+\pi^-$)
for a broader range of pseudorapidities including not only
main tracker but also calorimeters.
The problem of PID in the calorimeter is not clear to us. 

\section{Conclusions}

In this paper we have presented first predictions for the production
of two pairs of $e^+ e^-$ in ultraperipheral collisions
for leptons with
transverse momenta larger than some fixed values
characteristic for specific detectors at the LHC \cite{ALICE_epem}.
We have presented results for the full range of rapidities as well as 
the results taking into account experimental cuts on rapidities characteristic
for different experiments.

Before presenting our results for $e^+e^-e^+e^-$ production 
we have checked whether our approach describes the production
of a single $e^+e^-$ pair. A good agreement with the ALICE 
invariant mass distribution has been obtained.

Even imposing the experimental cuts relevant for different experiments 
we obtain cross sections that could be measured at the LHC even 
with relatively low luminosity required for UPC of heavy ions of the order of 
1 nb$^{-1}$. For instance, assuming the integrated luminosity of 1 nb$^{-1}$ 
for the main ATLAS detector angular coverage and transverse momentum cut
on each electron/positron
$p_t >$ 0.5 GeV we predict 235 events.

Measurements of two electrons of the same sign would be already a clear
signal of the double scattering mechanism.
In addition, one could measure also two dimensional distributions or
distributions in rapidity distance between two out of four produced
electrons. The electron and positron from the same scattering have well balanced
transverse momenta. They are also back-to-back in azimuthal angle.
Excluding such cases by imposing exclusion cuts in 
transverse momentum balance and/or azimuthal angle,
one could measure in coincidence electrons/positrons from different scatterings.

The distribution in relative azimuthal
angle between two electrons or two positrons is another interesting observable. 
Assuming dominance
of double scattering mechanism such a distribution should be flat
(constant when assuming no azimuthal correlation in lepton production
with respect to the nuclear scattering plane).  
Another, presented here, possibility is to measure 
distribution in relative rapidity distance between the same-sign and 
opposite-sign electrons. One could also measure corresponding invariant 
mass distributions (not discussed here) that are more difficult to
calculate, however, from purely technical reasons.

In future, for exact comparison to the measured cross sections 
a calculation of the single scattering 
$\gamma \gamma \to e^+ e^- e^+ e^-$ contribution may be also necessary.
This computation goes, however, beyond the scope of the present study, 
where we have concentrated exclusively on double scattering mechanism. 
We leave such a study for a future.

In summary, our analysis shows that first measurement(s) of the double
scattering in the $e^+ e^- e^+ e^-$ channel should be feasible.
We expect therefore a clear response to our proposal of all experimental groups
at the LHC.

\vspace{1.5cm}

{\bf Acknowledgments}

We are indebted to Janusz Chwastowski and Rafa{\l} Staszewski
for a discussion on possibilities of measuring the discussed here
double scattering processes by the ATLAS Collaboration.
This work was partially supported by the Polish grant 
No. DEC-2014/15/B/ST2/02528 (OPUS)
as well as by the Centre for Innovation and Transfer of Natural Sciences
and Engineering Knowledge in Rzesz\'ow.


\end{document}